\documentclass[12pt]{article}
\usepackage{graphicx}
\usepackage{subfigure}

\def\beq{\begin{equation}}
\def\eeq#1{\label{#1}\end{equation}}
\def\eeqn{\end{equation}}

\def\beqa{\begin{eqnarray}}
\def\eeqa#1{\label{#1}\end{eqnarray}}
\def\eeqan{\end{eqnarray}}

\let\bar=\overbar

\def\Dslash{\not{\hbox{\kern-4pt $D$}}}
\def\dslash{\not{\hbox{\kern-2pt $\del$}}}

\def\msb{{\bar{\ssstyle M \kern -1pt S}}}

\def\Title#1{\begin{center} {\Large {\bf #1} } \end{center}}

\begin{document}

\Title{\boldmath{$B\rightarrow l\nu$} - Belle results and outlook for Belle II}

\bigskip\bigskip

%+\addtocontents{toc}{{\it D. Reggiano}}
%+\label{ReggianoStart}

\begin{raggedright}  

{\it Yasuyuki Horii\\
Kobayashi-Maskawa Institute\\
Nagoya University\\
Furo-cho, Chikusa-ku, Nagoya, 464-8602, JAPAN}
\bigskip\bigskip
\end{raggedright}

\section{Introduction}

The leptonic decays $B^-\rightarrow \ell^- \bar{\nu}_\ell$
($\ell = e$, $\mu$, $\tau$) provide
opportunities for testing the Standard Model (SM) and for searching for new physics.\footnote{
Charge-conjugate decays are implied throughout this report unless otherwise stated.}
In the SM, the branching ratio ${\cal B}(B^{-}\rightarrow \ell^{-}\bar{\nu}_\ell)$
is proportional to $f_B^2|V_{ub}|^2$,
where $f_B$ is the $B$ meson decay constant 
and $V_{ub}$ is the Cabibbo-Kobayashi-Maskawa (CKM) matrix element~\cite{CKM}.
The expected branching ratios in the SM are $O(10^{-4})$,
$O(10^{-7})$, and $O(10^{-11})$ for $\ell = \tau$, $\mu$, and $e$, respectively,
where the mass difference between $\tau$, $\mu$, and $e$ affects the values.
Physics beyond the SM could suppress
or enhance ${\cal B}(B^{-}\rightarrow\ell^{-}\bar{\nu}_\ell)$ 
via exchange of a new charged particle such as a charged Higgs boson~\cite{TypeII, RPV, NUHM2}.
In this report, the results obtained at the Belle experiment are introduced.
An outlook for the Belle II experiment is also shown.

\section{Belle results}

The leptonic decays $B^-\rightarrow \ell^- \bar{\nu}_\ell$ include neutrinos in the final state,
which cannot be detected by the Belle detector.
At the Belle experiment, it is exploited that a $B$ meson pair is generated
from the process $e^+e^-\rightarrow \Upsilon(4S)\rightarrow B\bar{B}$.
We reconstruct one of the $B$ mesons~(``$B_{\rm tag}$'')
and identify the signal decays in the other $B$ mesons~(``$B_{\rm sig}$'').
%The neutrinos are identified by using the extra energy in the calorimeter
%after removing the reconstructed particles and the missing mass squared in the event.

\begin{figure}[htbp]
 \begin{center}
  \leavevmode
  \subfigure[The extra energy $E_{\rm ECL}$ distribution in Ref.~\cite{Belle_Semi_2010}.]
  {\includegraphics[width=0.3\textwidth]{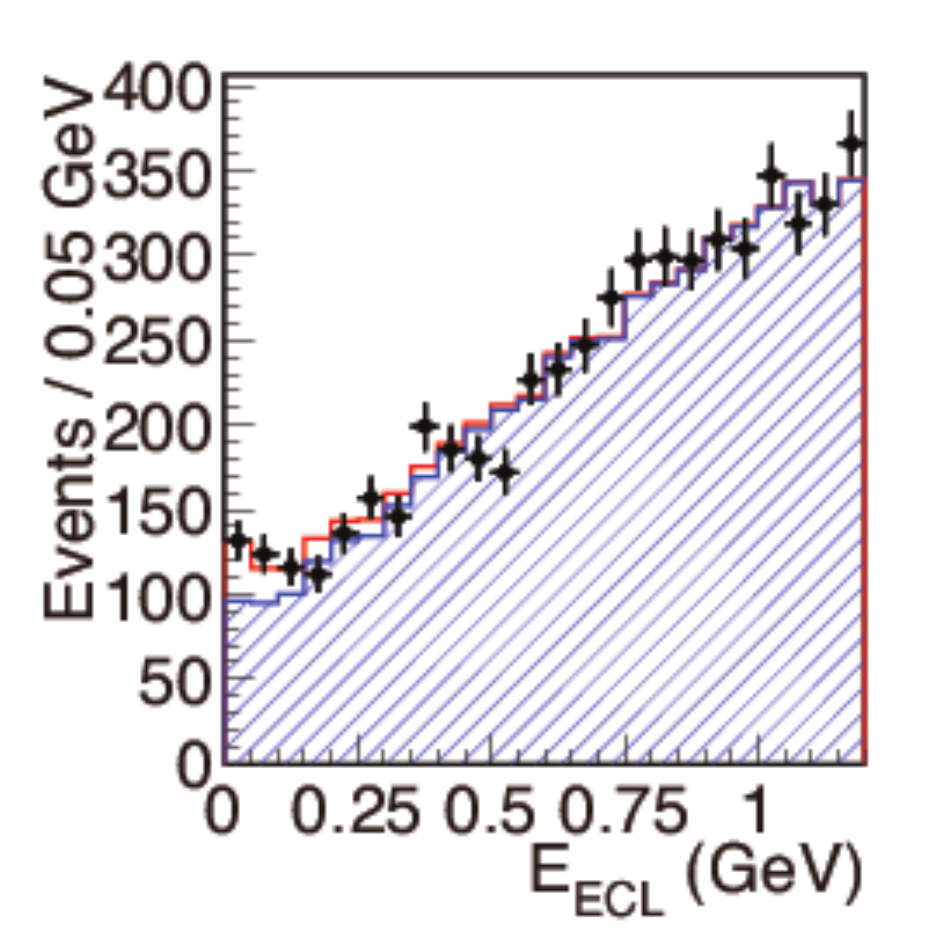}}
  \hspace{2mm}
  \subfigure[The extra energy $E_{\rm ECL}$ distribution in Ref.~\cite{Belle_Had_2012}.]
  {\includegraphics[width=0.31\textwidth]{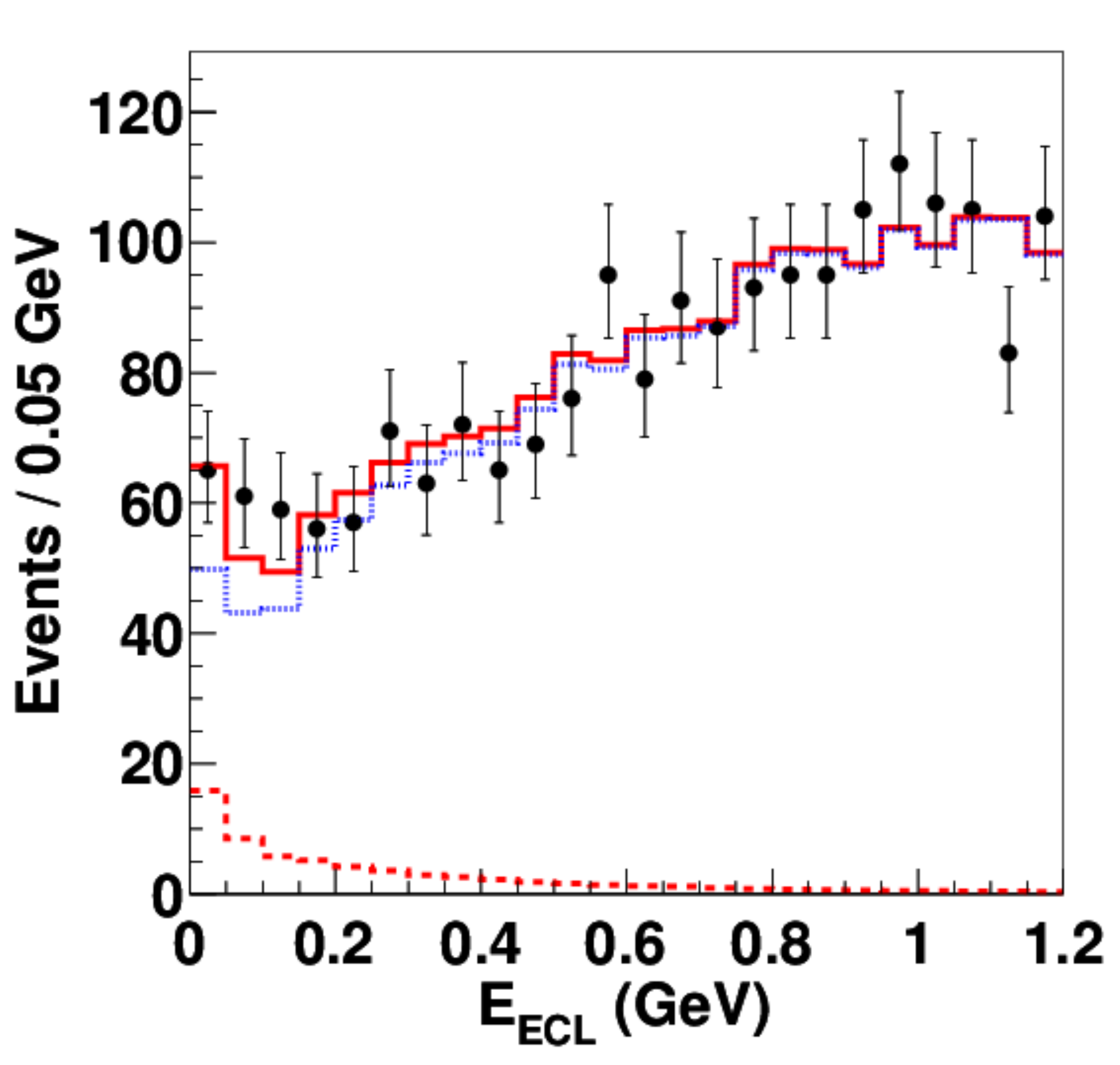}}
  \hspace{0mm}
  \subfigure[The missing mass squared $M_{\rm miss}^2$ distribution
  ($E_{\rm ECL} < 0.2~{\rm GeV}$) in Ref.~\cite{Belle_Had_2012}.]
  {\includegraphics[width=0.31\textwidth]{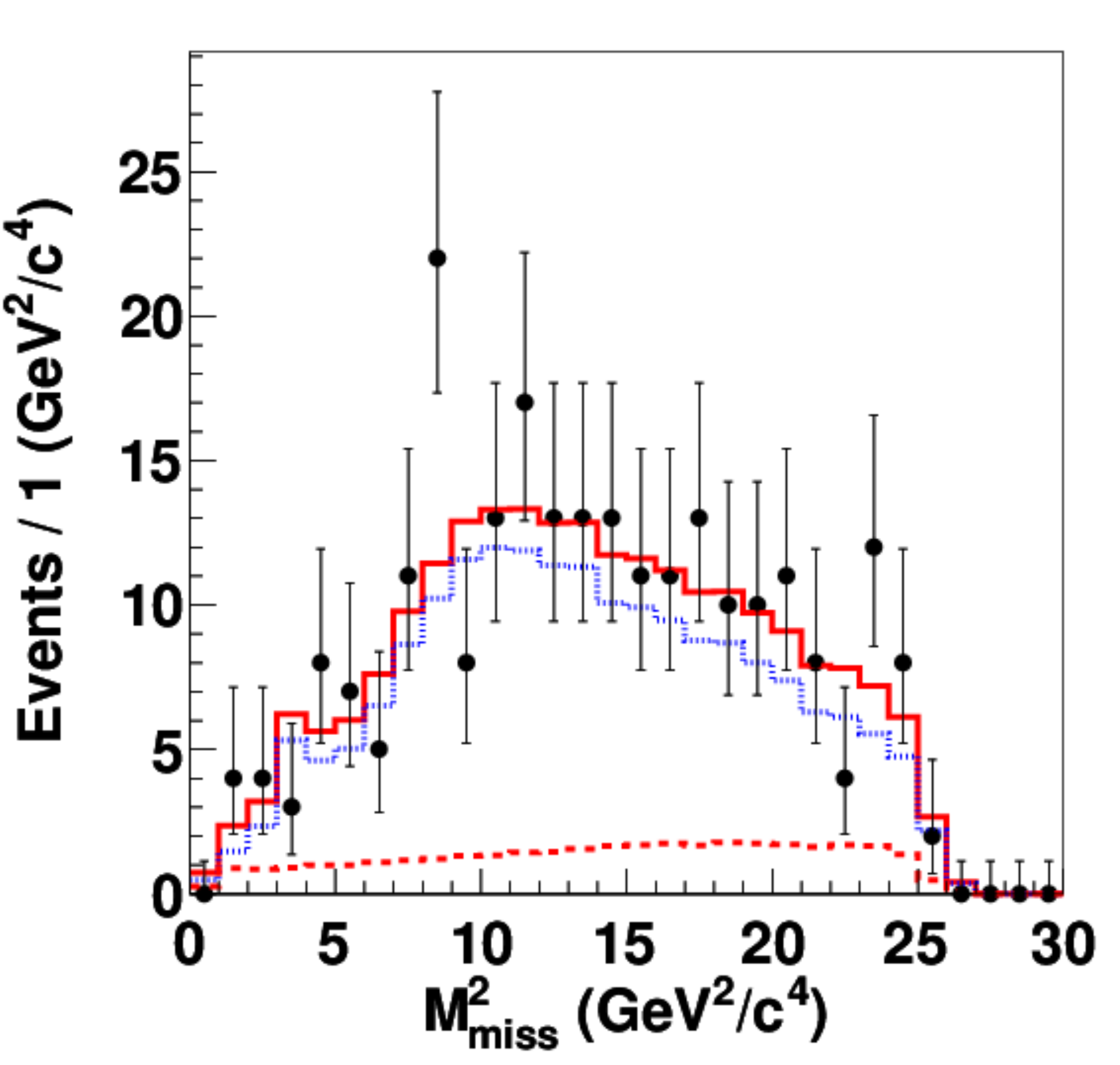}}
  \caption{Signal extraction for $B^-\rightarrow \tau^-\bar{\nu}_\tau$ at the Belle experiment.
  The solid circles with error bars are data.
  The solid histograms show the fit results.
  The dashed histograms in (b) and (c) show the signal component,
  while the hatched histogram in (a) and the dotted histograms in (b) and (c) show the background component.}
  \label{fig:belle_result}
 \end{center}
\end{figure}

The first evidence of $B^{-}\rightarrow\tau^{-}\bar{\nu}_\tau$
was reported by the Belle collaboration
with a significance of 3.5 standard deviations ($\sigma$)
%including systematic uncertainty
and a measured branching ratio of 
${\cal B}(B^{-}\rightarrow\tau^{-}\bar{\nu}_\tau)
= [1.79^{+0.56}_{-0.49}(\mbox{stat})^{+0.46}_{-0.51}(\mbox{syst})] \times 10^{-4}$~\cite{Belle_Had_2006}.
This measurement used hadronic modes for reconstructing $B_{\rm tag}$ (``hadronic tag'')
and a data sample corresponding to $449 \times 10^6$ $B\bar{B}$ events.
This was followed by a measurement of ${\cal B}(B^{-}\rightarrow\tau^{-}\bar{\nu}_\tau)
= [1.54^{+0.38}_{-0.37}({\rm stat})^{+0.29}_{-0.31}({\rm syst})]\times 10^{-4}$
with a significance of 3.6$\sigma$ (Fig.~\ref{fig:belle_result}~(a))~\cite{Belle_Semi_2010}.
%including systematic uncertainty (Figure~\ref{fig:belle_result}~(a))~\cite{Belle_Semi_2010}.
This measurement used semileptonic  modes
for reconstructing $B_{\rm tag}$ (``semileptonic tag'')
and a data sample corresponding to $657 \times 10^6$ $B\bar{B}$ events.
%Both results are slightly higher than the estimate $(0.73^{+0.12}_{-0.07})\times 10^{-4}$
%based on a global fit to the CKM matrix elements assuming the SM~\cite{CKMfitter}.
%Combining with the results obtained from the BaBar collaboration,
%a world average was obtained to be $(1.67 \pm 0.30) \times 10^{-4}$~\cite{HFAG}.
%This result was nearly $3\sigma$ higher than the estimate $(0.73^{+0.12}_{-0.07})\times 10^{-4}$
%based on a global fit to the Cabibbo-Kobayashi-Maskawa (CKM) matrix elements~\cite{CKMfitter}.

\begin{figure}[htbp]
\begin{center}
  \leavevmode
  \subfigure[Comparison for ${\cal B}(B^{-}\rightarrow\tau^{-}\overline{\nu}_{\tau})$
  between the direct measurement, the Belle result as well as
  a world average ${\cal B}(B^{-}\rightarrow\tau^{-}\overline{\nu}_{\tau}) = (1.15\pm 0.23)\times 10^{-4}$,
  and a SM estimate from the CKM global fit~\cite{CKMfitter}.]
  %For the direct measurement, we show the Belle result as well as a world average of $(1.15\pm 0.23)\times 10^{-4}$.]
  {\includegraphics[width=0.45\textwidth]{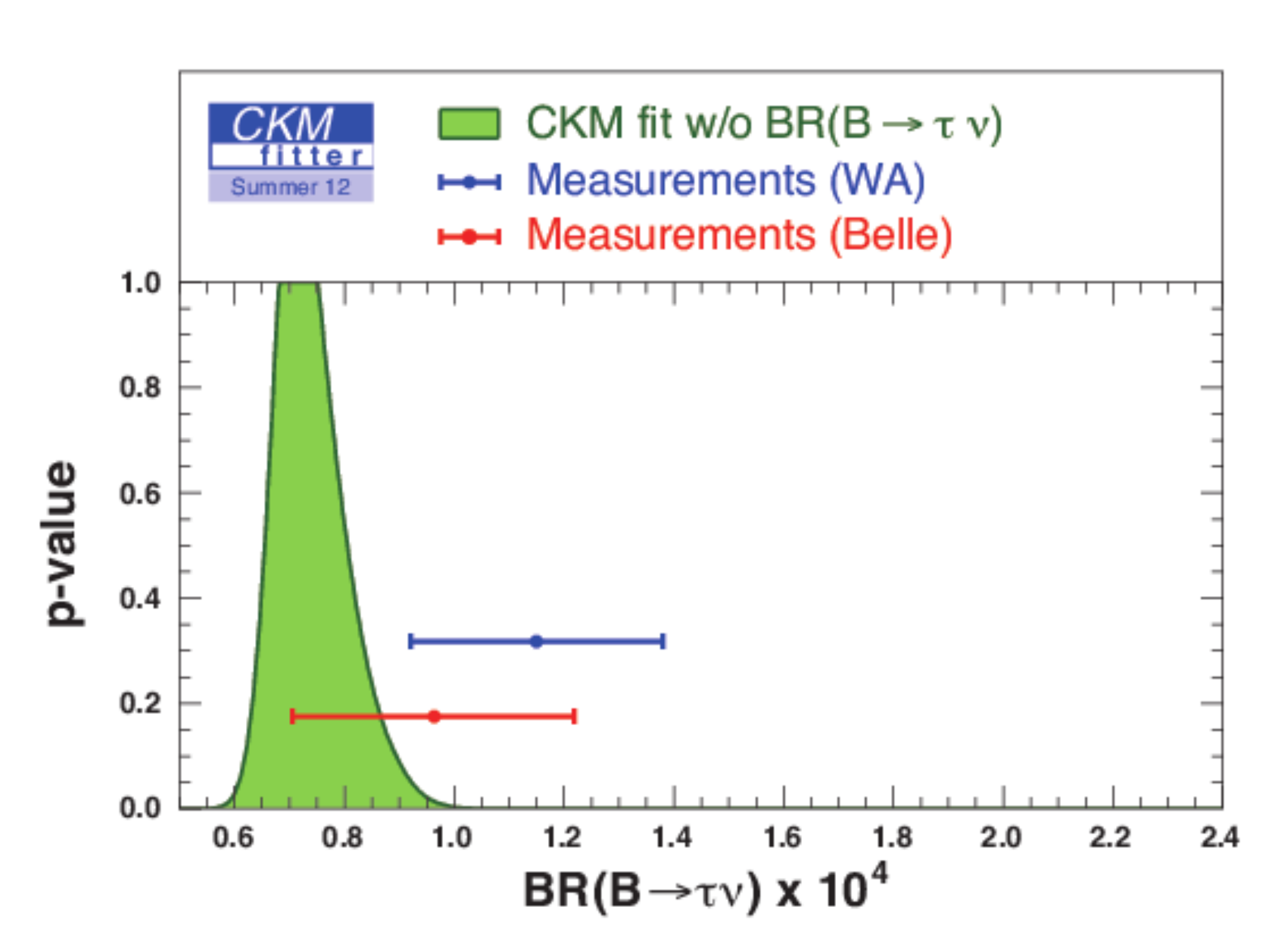}}
  \hspace{6mm}
  \subfigure[Constraint on $\tan\beta$ %the ratio of the two vacuum expectation values
  and the charged Higgs mass in the type II two Higgs doublet model~\cite{TypeII}.
  The filled regions indicate excluded regions at 95\% C.L.
  The factors $f_B = (190 \pm 9)$~MeV~\cite{HPQCD}
  and $|V_{ub}| = (4.15 \pm 0.49) \times 10^{-3}$~\cite{PDG} are used.]
  {\includegraphics[width=0.4\textwidth]{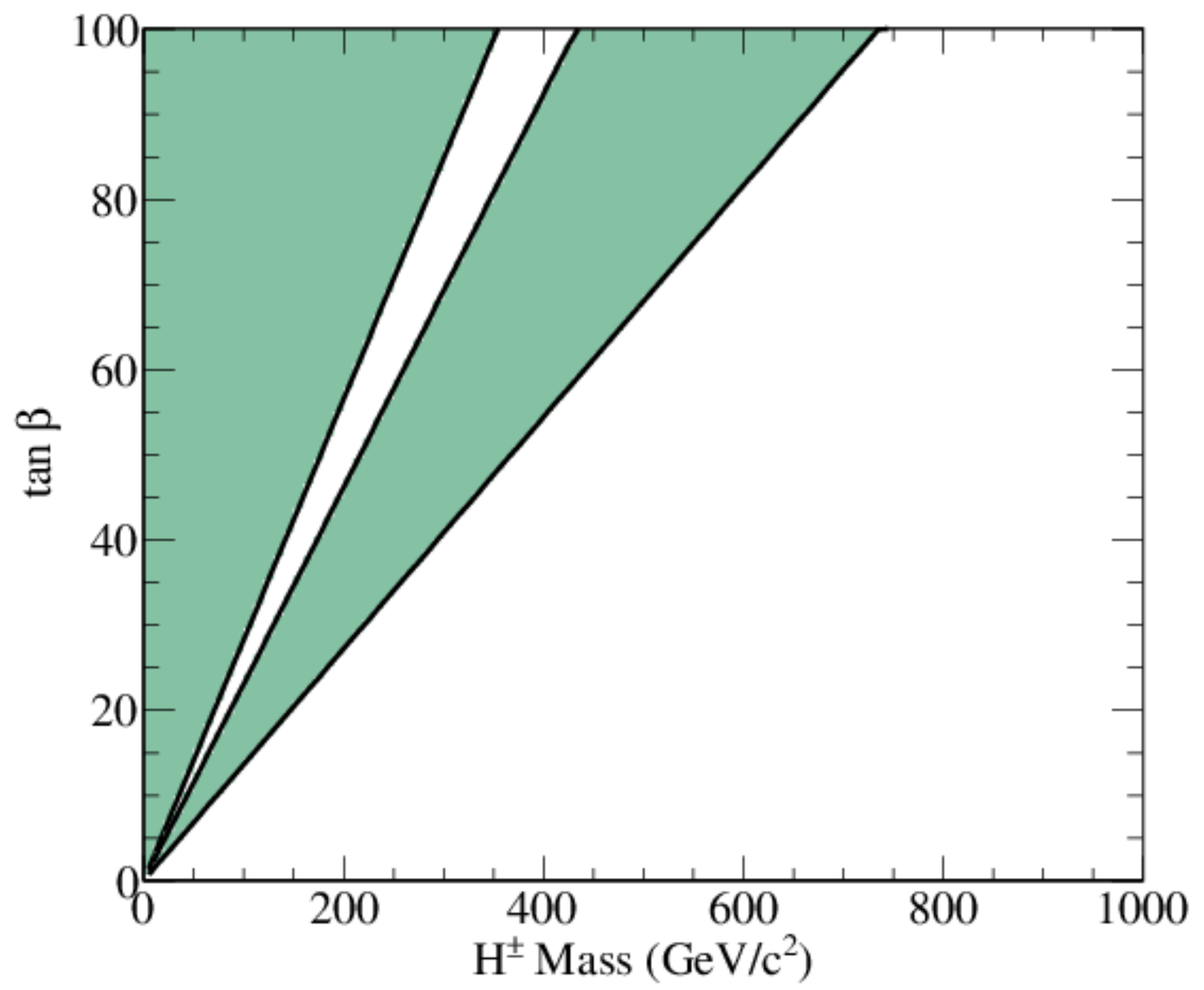}}
    \caption{A comparison of the ${\cal B}(B^{-}\rightarrow\tau^{-}\overline{\nu}_{\tau})$
    result with a SM estimate %based on the CKM global fit
    and a constraint on the type II two Higgs doublet model.
    We use the Belle's combined result of ${\cal B}(B^{-}\rightarrow\tau^{-}\overline{\nu}_{\tau}) = (0.96\pm 0.26)\times 10^{-4}$.}
    \label{fig:constraint}
\end{center}
\end{figure}

In the summer of 2012, the Belle collaboration updated the result for the hadronic tag
using Belle's final data sample corresponding to $772 \times 10^6$ $B\bar{B}$ events~\cite{Belle_Had_2012}.
The branching ratio is obtained to be ${\cal B}(B^{-}\rightarrow\tau^{-}\bar{\nu}_\tau)
= [0.72^{+0.27}_{-0.25}({\rm stat})\pm 0.11({\rm syst})]\times 10^{-4}$
with a significance of $3.0\sigma$
%including systematic uncertainty
(Fig.~\ref{fig:belle_result}~(b) and (c)).
By employing a neural network-based method for the hadronic tag
and a two-dimensional fit for the signal extraction, along with a larger data sample,
both statistical and systematic precisions are significantly improved.
Combined with the measurement
based on the semileptonic tag~\cite{Belle_Semi_2010}
taking into account all the correlated systematic errors,
the branching ratio is found to be 
${\cal B}(B^{-}\rightarrow\tau^{-}\overline{\nu}_{\tau}) =
(0.96 \pm 0.26) \times 10^{-4}$
with a significance of $4.0\sigma$.
%including systematic uncertainties.
The result is consistent with the value $(0.72^{+0.12}_{-0.08})\times 10^{-4}$
obtained from a global fit to CKM matrix elements assuming the SM (Fig.~\ref{fig:constraint}~(a))~\cite{CKMfitter}.
%This value is consistent with the SM expectation
%obtained from the CKM global fit (Figure~2~(a)).
Using our new combined result and parameters found in Ref.~\cite{PDG}, we obtain
$f_B |V_{ub}|= [7.4 \pm 0.8({\rm stat}) \pm 0.5({\rm syst})] \times 10^{-4}~{\rm GeV}$.
Our result also provides constraints on new physics models
including charged Higgs bosons (Fig.~\ref{fig:constraint}~(b)).

For the $B^{-}\rightarrow \mu^{-}\bar{\nu}_\mu$ and $B^{-}\rightarrow e^{-}\bar{\nu}_e$ decays,
significant signal has not been obtained.
The Belle collaboration obtained the upper limits of
${\cal B}(B^{-}\rightarrow \mu^{-}\bar{\nu}_\mu) < 1.7 \times 10^{-6}$ and
${\cal B}(B^{-}\rightarrow e^{-}\bar{\nu}_e) < 9.8 \times 10^{-7}$
at 90\% confidence level (C.L.)
by reconstructing $B_{\rm tag}$ using all the remaining particles
after removing $\mu^-$ and $e^-$ in the $B_{\rm sig}$ reconstruction (``inclusive tag'')~\cite{Belle_lnu_inclusive}.
This measurement used a data sample corresponding to $657 \times 10^6$ $B\bar{B}$ events.
In the summer of 2012, the Belle collaboration also obtained the upper limits of
${\cal B}(B^{-}\rightarrow \mu^{-}\bar{\nu}_\mu) < 2.5 \times 10^{-6}$ and
${\cal B}(B^{-}\rightarrow e^{-}\bar{\nu}_e) < 3.5 \times 10^{-6}$ at 90\% C.L.
using the hadronic tag and the full data sample~\cite{Belle_lnu_hadronic}.
The results are consistent with the SM expectations.
%of ${\cal O}(10^{-7})$ and ${\cal O}(10^{-11})$ for $\mu^-$ and $e^-$ modes, respectively.

\section{Outlook for Belle II}

The KEKB collider and the Belle detector %used for the Belle experiment
will be upgraded to the SuperKEKB collider and the Belle II detector, respectively~\cite{TDR}.
The design center-of-mass energy for SuperKEKB
is on the $\Upsilon(4S)$ resonance, which is the same as KEKB.
The design luminosity for SuperKEKB
is $8.0\times 10^{35}~{\rm cm}^{-2}{\rm s}^{-1}$,
which is about 40 times larger than the current world record
of $2.1\times 10^{34}~{\rm cm}^{-2}{\rm s}^{-1}$ achieved by KEKB.
The target integrated luminosity is 50 ab$^{-1}$, which is about 50 times larger than KEKB.
%The main concern for the Belle II detector is the higher-background environment.
The Belle~II detector is designed to cope with higher background,
which is estimated by extrapolating the results of the operations of KEKB and Belle.
The Belle~II detector has better performance than the Belle detector
in the vertex determination based on the pixel detectors
and the particle identification based on the imaging Cherenkov detectors.
The physics run is planned to start in 2016.
%For evaluating the effects of the backgrounds on the detector performance,
%we extrapolate the results of the operations of KEKB and Belle
%by accounting for scaling for each component of backgrounds.
%The Belle II detector is designed to cope with this estimation conservatively,
%and has better performance than the Belle detector
%in vertex determination and particle identification.

%At the SuperKEKB collider and the Belle II detector,
%we have broad physics program in the fields of heavy flavor physics.
%Our measurements will over-constrain the parameter space
%of the Standard Model and its extensions and will shed light on the nature of new physics
%by exploiting the correlations among various observables.
The measurements of ${\cal B}(B^-\rightarrow \ell^- \bar{\nu}_\ell)$
are one of the most important physics programs at the Belle II experiment.
A key experimental issue is to understand
the extra energy distribution (see Fig.~\ref{fig:belle_result} (a) and (b)),
which is a main variable for discriminating the signal from the background,
over the higher beam background level.
It is also important to improve the theoretical understanding of $f_B$
and the precision of  $|V_{ub}|$ determination from the $b\rightarrow u$ transitions
so that we can use ${\cal B}(B^-\rightarrow \ell^- \bar{\nu}_\ell)$ for a test of SM and new physics.

\begin{figure}[htbp]
\begin{center}
  \leavevmode
  \subfigure[Constraint on $\tan\beta$
  and the charged Higgs mass in the type II two Higgs doublet model~\cite{TypeII}.
  The filled regions indicate excluded regions at 95\% C.L.
  Assumed central values are ${\cal B}(B^-\rightarrow \tau^- \bar{\nu}_\tau) = 1 \times 10^{-4}$,
  $f_B = 190$~MeV, and $|V_{ub}| = 4.15 \times 10^{-3}$.
  Assumed error for $f_B^2 |V_{ub}|^2$ is relatively~4\%.]
  {\includegraphics[width=0.4\textwidth]{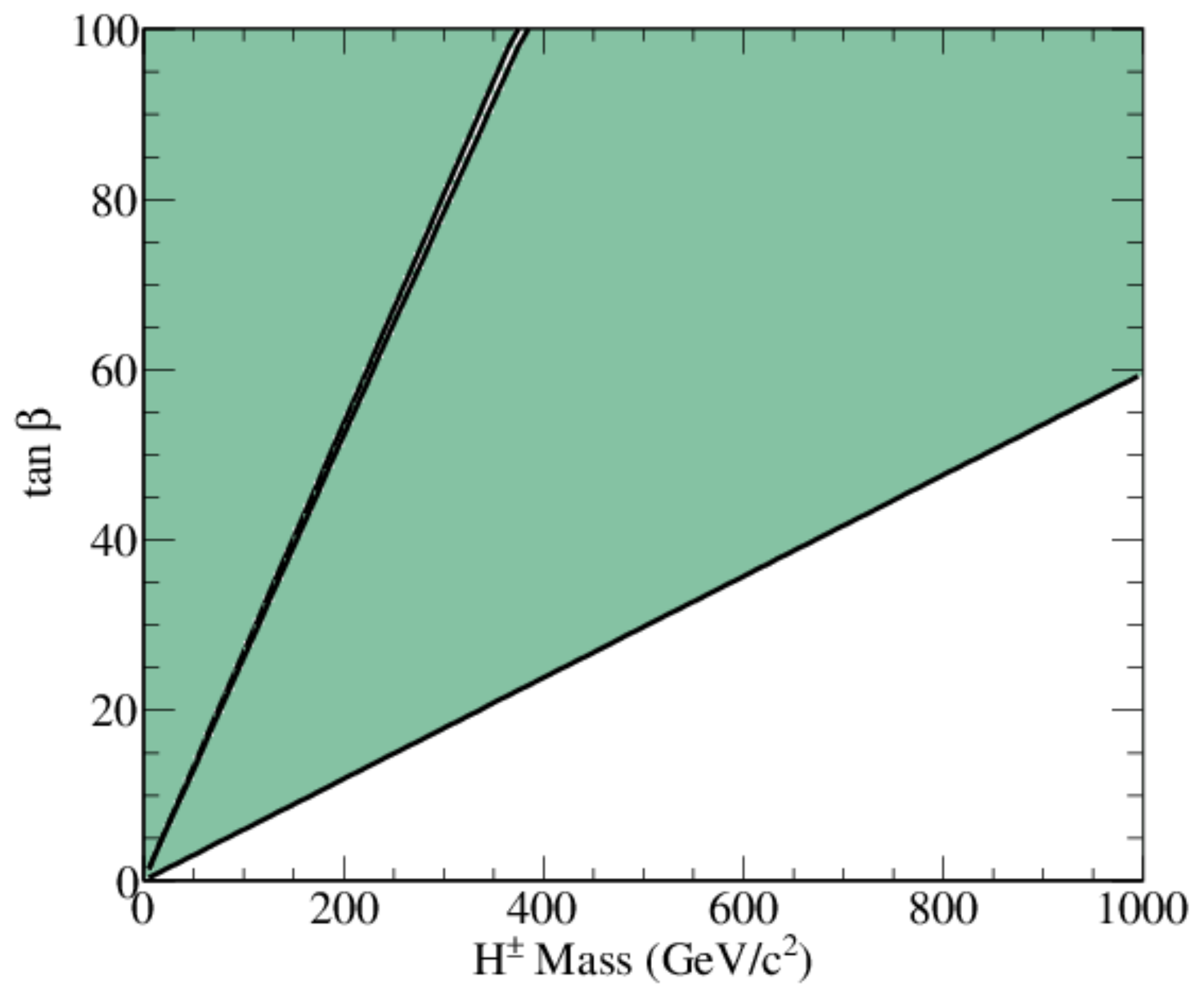}}
  \hspace{8mm}
  \subfigure[Constraint on the 2-parameter nonuniversal Higgs model~\cite{NUHM2}.
  The blue and orange dots are obtained for allowed parameter regions
  in $m_0 < 5$~TeV and $m_0 < 20$~TeV, respectively.
  The black dotted and solid lines show the $\pm 2\sigma$ boundary and the central value,
  respectively, for the world average of ${\cal B}(B^-\rightarrow \tau^- \bar{\nu}_\tau)$ in 2007.
  The green region shows an expectation for Belle~II.]
  {\includegraphics[width=0.45\textwidth, bb = 0 0 580 390]{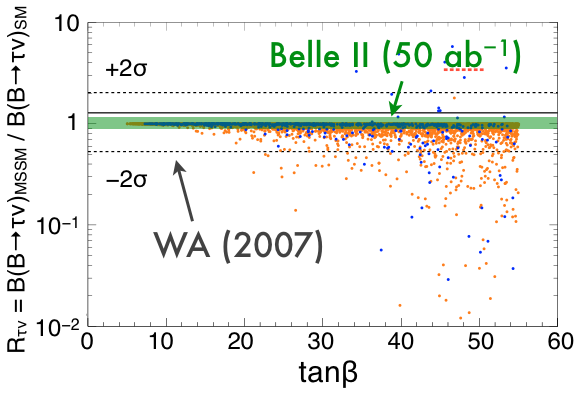}}
    \caption{Expectation for the constraint on new physics models
    at an integrated luminosity of 50 ab$^{-1}$ at Belle II.}
    \label{fig:outlook}
\end{center}
\end{figure}

For ${\cal B}(B^-\rightarrow \tau^- \bar{\nu}_\tau)$, we expect a precision
of about 10\% and a few \% at the integrated luminosity of 5 ab$^{-1}$ and 50 ab$^{-1}$, respectively.
For this measurement, we need to estimate the systematic uncertainties with better precision.
The main systematic uncertainties were estimated by using the data sample
at the Belle experiment, and a meaningful improvement is expected
using increased data sample at the Belle II experiment.
Fig.~\ref{fig:outlook} shows the expectations
for the constraint on the type II two Higgs doublet model~\cite{TypeII}
and the 2-parameter nonuniversal Higgs model with a Higgs boson mass of 125 GeV~\cite{NUHM2}.
Stringent constraints on these models are expected to be obtained. %from ${\cal B}(B^-\rightarrow \tau^- \bar{\nu}_\tau)$. %at the Belle II experiment.
For ${\cal B}(B^-\rightarrow \mu^- \bar{\nu}_\mu)$,
we expect a 5$\sigma$ observation at an integrated luminosity of 5~ab$^{-1}$
assuming the SM branching ratio.
For both ${\cal B}(B^-\rightarrow \mu^- \bar{\nu}_\mu)$ and ${\cal B}(B^-\rightarrow e^- \bar{\nu}_e)$,
we expect a sensitivity of ${\cal O}(10^{-8})$ at an integrated luminosity of 50~ab$^{-1}$.
These measurements will provide important insight into lepton universality.

\section{Conclusion}

The leptonic decays $B^-\rightarrow \ell^- \bar{\nu}_\ell$ provide
opportunities for testing the SM and for searching for new physics.
Evidence of signal has been obtained for $B^-\rightarrow \tau^- \bar{\nu}_\tau$
using hadronic and semileptonic tags at Belle.
Upper limits have been obtained
for $B^-\rightarrow \mu^- \bar{\nu}_\mu$ and $B^-\rightarrow e^- \bar{\nu}_e$
using inclusive and hadronic tags at Belle.
All the results are consistent with the SM,
and provide constraints on new physics models
including charged Higgs bosons.
Stringent tests of the SM and new physics
will be performed by the measurements with better precision at Belle II.

%%%%%%%%%%%%%%%%%%%%%%%%%%%%%%%%%%%%%%%%%%%%%%%%%%%%%%%%%%%%%%%%%%%%%%%%%
%%
%%   use this format to include an .eps figure into your paper
%%
\if0
\begin{figure}[htb]
\begin{center}
\epsfig{file=magnet.eps,height=1.5in}
\caption{Plan of the magnet used in the Mesmeric studies.}
\label{fig:magnet}
\end{center}
\end{figure}
\fi
%%%%%%%%%%%%%%%%%%%%%%%%%%%%%%%%%%%%%%%%%%%%%%%%%%%%%%%%%%%%%%%%%%%%%%%%%%%

%%%%%%%%%%%%%%%%%%%%%%%%%%%%%%%%%%%%%%%%%%%%%%%%%%%%%%%%%%%%%%%%%%%%%%%%%
%%
%%   use this format to include a LaTeX table  into your paper
%%
\if0
\begin{table}[b]
\begin{center}
\begin{tabular}{l|ccc}  
Patient &  Initial level($\mu$g/cc) &  w. Magnet &  
w. Magnet and Sound \\ \hline
 Guglielmo B.  &   0.12     &     0.10      &     0.001  \\
 Ferrando di N. &  0.15     &     0.11      &  $< 0.0005$ \\ \hline
\end{tabular}
\caption{Blood cyanide levels for the two patients.}
\label{tab:blood}
\end{center}
\end{table}
\fi
%%%%%%%%%%%%%%%%%%%%%%%%%%%%%%%%%%%%%%%%%%%%%%%%%%%%%%%%%%%%%%%%%%%%%%%%%%%

\bigskip

This presentation is supported by JSPS KAKENHI Grant Number 24740157.

\end{document}